\documentclass[conference]{IEEEtran}
\IEEEoverridecommandlockouts
\usepackage{cite}
\usepackage{amsmath,amssymb,amsfonts}
\usepackage{algorithmic}
\usepackage{graphicx}
\usepackage{textcomp}
\usepackage{xcolor}
\usepackage{booktabs}
\usepackage{siunitx}
\usepackage{tabularx}
\usepackage{listings}
\usepackage{subcaption}

\usepackage{threeparttable}
\usepackage{enumitem}
\usepackage{hyperref}

\def\BibTeX{{\rm B\kern-.05em{\sc i\kern-.025em b}\kern-.08em
    T\kern-.1667em\lower.7ex\hbox{E}\kern-.125emX}}
\begin{document}

\title{Attacks Against Mobility Prediction in 5G Networks
\thanks{This work is supported by framework grant RIT17-0032 from the Swedish Foundation for Strategic Research (SSF).}
}

\author{\IEEEauthorblockN{Syafiq Al Atiiq, Yachao Yuan, Christian Gehrmann}
\IEEEauthorblockA{
\textit{Lund University}\\
Lund, Sweden \\
\{syafiq\_al.atiiq, yachao.yuan, christian.gehrmann\}@eit.lth.se}
\and
\IEEEauthorblockN{Jakob Sternby, Luis Barriga}
\IEEEauthorblockA{\textit{Ericsson Research} \\
Lund \& Stockholm, Sweden \\
\{jakob.sternby, luis.barriga\}@ericsson.com}
}

\maketitle

\begin{abstract}
The $5^{th}$ generation of mobile networks introduces a new Network Function (NF) that was not present in previous generations, namely the Network Data Analytics Function (NWDAF). Its primary objective is to provide advanced analytics services to various entities within the network and also towards external application services in the 5G ecosystem. One of the key use cases of NWDAF is mobility trajectory prediction, which aims to accurately support efficient mobility management of User Equipment (UE) in the network by allocating ``just in time'' necessary network resources. In this paper, we show that there are potential mobility attacks that can compromise the accuracy of these predictions. In a semi-realistic scenario with 10,000 subscribers, we demonstrate that an adversary equipped with the ability to hijack cellular mobile devices and clone them can significantly reduce the prediction accuracy from 75\% to 40\% using just 100 adversarial UEs. While a defense mechanism largely depends on the attack and the mobility types in a particular area, we prove that a basic KMeans clustering is effective in distinguishing legitimate and adversarial UEs. 
\end{abstract}

\begin{IEEEkeywords}
NWDAF, 5G, mobility prediction, adversarial mobility
\end{IEEEkeywords}

\section{Introduction}

According to the recent Ericsson mobility report\footnote{https://www.ericsson.com/en/reports-and-papers/mobility-report}, 5G mobile networks will reach 5 billion subscriptions by 2028. Compared to the previous generations, 5G will become the main subscription type in the coming years due to its support for higher speeds, higher bandwidth, lower latencies, improved security/privacy, and network capabilities exposure towards industry verticals and enterprises. For this reason, 5G is also considered a platform that enables innovation to investigate novel data-driven use cases, applications, and automation. NWDAF is a new network function that the 3G Partnership Project (3GPP) formalized in response to the maturity of Artificial Intelligence (AI) in recent years. NWDAF intends to benefit from the explosion of data generated by 5G. Its architecture and API are documented in 3GPP Technical Specification (TS) 23.288 \cite{3gpp.23.288} and TS 29.520 \cite{3gpp.29.520}, respectively. NWDAF's main objective is to provide internal analytics service to other NFs to automate processes, i.e., troubleshooting, resource allocation, and anomaly detection.

Based on \cite{3gpp.23.288,3gpp.29.520}, one of the use cases of NWDAF is UE mobility analytics and prediction. For this context, the prediction can be used to optimize the resource allocation to be as precise as possible based on the user's needs. This is to enhance user experience and efficiency, as edge computing resources for heavy applications need to be pre-allocated before the user moves into a particular area. TS 37.817 \cite{3gpp.37.817} mentioned that trajectory prediction can be used to reduce late/early or wrong handovers. Looking from the opposite angle, the same mobility prediction can be used to detect whether a particular UE mobility is possibly adversarial. The definition of adversarial UE mobility itself is broad, but we are particularly interested in the deliberate actions of UEs targeting the accuracy of a mobility model within NWDAF. Therefore, in our context, an adversarial UE is one that tries to fool such a model and ultimately degrade its utility.

This paper investigates possible attacks against a mobility trajectory model in NWDAF. In a real-world scenario, the model may be implemented differently, depending on the UE's mobility type, coverage, and density in a particular area. Given a set of legitimate UE movements, we would like to understand how the malicious activity of adversarial UEs could influence the prediction of the mobility model implemented in NWDAF. Since NWDAF is a new concept in 3GPP, attacks against its mobility trajectory models are yet to be found. While some of our attacks are hypothetical, based on the assumption that the adversary has control over a number of UEs, others are based on similar, earlier attempted attacks against the existing system. As a presumption, our work does not consider outliers in the network, i.e., certain events that gather large crowds. We experimentally evaluated the best model to be used for legitimate mobility, then attacked the same model with the intention of decreasing its prediction accuracy. To the best of our knowledge, this is the first work to investigate and evaluate attacks against mobility prediction models in NWDAF. As a follow-up, we investigate how to differentiate the adversarial UE movements from the legitimate UEs. This can be used to prevent the mobile operator from including the adversarial UEs in the next retraining period so that the model degradation quality can be avoided. 

The paper is organized as follows. We overview the problem definition in Section \ref{sec:problem}. We manifest the UE mobility in Section \ref{sec:mobility}. The simulation environment and the results are written in sections \ref{sec:simulation} and \ref{sec:result}, respectively. Defense mechanisms for the respective attacks are discussed in \ref{sec:def}. We present the related work in section \ref{sec:related} and finally draw the conclusion in section \ref{sec:conclusion}.

\section{Problem Definition}
\label{sec:problem}
NWDAF collects data streams from neighboring NFs and performs analysis on the streams to support NF's tasks through an NWDAF subscription model. In this paper, we investigate adversarial mobility patterns and how they can influence NWDAF mobility models, which in turn will impact operations for the NFs that utilize these models. The mobility data is usually streamed from the Mobility Management Entity - MME (for 4G) or Access and Mobility Management Function - AMF (for 5G). In this work, we do not differentiate mobility data by origin; therefore, the base station in the system, i.e., eNodeB (4G) and gNodeB (5G), will be used interchangeably. 

\subsection{Scenario}
\begin{figure}[ht]
    \centering
    \includegraphics[width=0.85\linewidth]{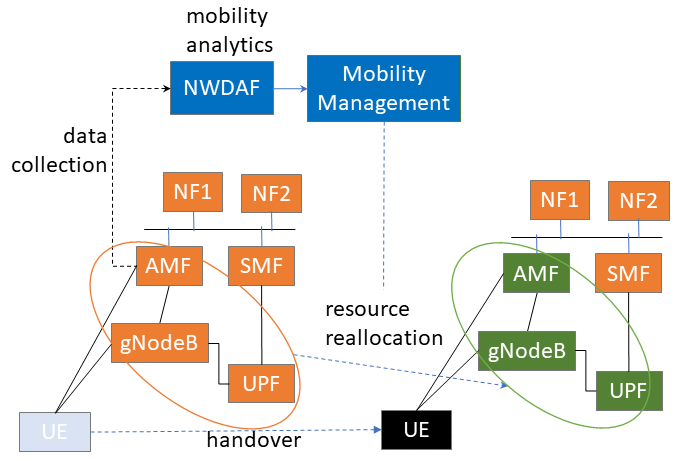}
    \caption{Use Case Scenario}
    \label{fig:scenario}
\end{figure}

We consider the NWDAF's use case of UE's mobility analysis and prediction. In our case, the output of mobility prediction is often consumed by the mobility management subsystem. It may also be used by its surrounding support services (for simplicity, let's call this an ``application server''). Mobility predictions are useful to prepare the necessary network resources before the UE moves into another location. In a non-malicious setting, this will enable smooth handover and provide optimal data traffic routing \cite{9392779}. Figure \ref{fig:scenario} depicts the scenario of a moving UE in a 5G setting. Initially, the UE is attached to some radio and Core Network (CN) NFs - gNodeB, Access and Mobility Management Function (AMF), Session Management Function (SMF), and User Plane Function (UPF) - commonly via the closest edge site.

At some point, the UE moves into a different geographical area where the currently allocated resources are sub-optimal. This move implies (i.) radio handover (changes of gNodeB), (ii) CN handover (AMF \& UPF) re-selection, and (iii) the change of the application server (not shown in the picture). With the introduction of mobility prediction in NWDAF, the aim is to perform this mechanism automatically such that the target network resources at the destination are ready for the handover when the UE is moving. Practically, this will smoothen the process and provide a more efficient handover. For high-mobility UEs, mobility management manages network resource re-allocation in order to ensure service continuity and QoS. This implies efficient handovers across radio and core network resources such as gNodeB, AMF, UPF, and Application Function (AF). Note that a radio handover between gNodeBs doesn't necessarily imply a handover between CN nodes such as UPF. That is dependent on the network configuration. Upon handovers, the network may need to spawn new resources on the destination and reduce the ones in the original location.

Adversarial UEs aware of this scenario could try to induce the mobility trajectory model deployed in NWDAF with malicious movement, i.e., moving around the network with the intent of decreasing the accuracy of the prediction performance of the model. With reduced prediction performance, network resources could be wrongly allocated, causing inefficiency and decreased Quality of Service (QoS).

\subsection{Threat Model}
\label{threadmodel}

We assume an adversary has the power to acquire UEs and then clone \cite{liu2015cloning,10.1007/978-3-319-24174-6_24,10.1007/978-3-030-68487-7_9,10.1145/3447993.3483254,p1sec} them to obtain a number of UEs with the same identity. Acquiring UE does not necessarily mean that the adversary needs to steal sim cards, but legitimately owning a sim card can also be counted \cite{gmaps-attack}. A practical example of this action is when a powerful adversary legally acquires a large amount of IoT devices in a short time. In this case, there is no need to hijack anything, and cloning the device is much easier as the source and cloned devices are already owned by the adversary. 

Similar types of attacks might increase considerably in the future. For example, the 3GPP security report TR 33.861 \cite{3gpp_tr_33_861} identified two threats stemming from the massive deployment of cellular IoT devices in 5G, stating that a malicious attacker could take over the same application on a large number of UEs and instruct them to launch signaling attacks at the same time. In our case, we argue that the UE application can be manipulated to generate malicious mobility patterns instead. The same document also mentions the risk of massive low-security unattended cellular IoT devices, such as shared bicycles, that can be hijacked and used to launch Denial of Service (DoS) attacks toward the radio network. In our case, we argue that the attack can be used to generate malicious mobility patterns instead. The Mirai botnet is an example of how the infection could also possibly spread across massive cellular IoT.

An adversary may increase the number of cloned UEs to many ``perceived to be'' same UEs by the operator. It has been shown that extracting the USIM (Universal Subscriber Identity Module) master secret key in 4G is feasible \cite{liu2015cloning}\cite{10.1007/978-3-319-24174-6_24}. A more advanced side channel attack to compromise USIM has also been shown using deep learning \cite{10.1007/978-3-030-68487-7_9}. Compromising the master USIM key means that an adversary has the capability of cloning them. Aside from compromising the USIM, a temporal UE's cloning can also be achieved by intercepting the authentication vectors generated by the network and stealing them via a false base station\cite{p1sec}. While this was proven to be successful against 2G, 3G, and 4G, the same issue will be faced by 5G interconnects if the operator does not enforce a coherency control on the inbound 5G signaling \cite{p1sec}.

In this work, we present several malicious mobilities, explained in more detail in section \ref{mob:adv}. The objective of such mobilities is to decrease the overall prediction accuracy of a mobility model. From the operator's perspective, a decreased accuracy would mean two things: (i.) This can lead the mobile operator to believe that there is a shift in UE mobility patterns, which may prompt them to retrain the model with newer data. (ii.) if the former is executed, then the prediction quality is significantly downgraded, which means the predictions of the model are unreliable. Consequently, the efficiency, as well as the resource allocation function of networks, would drastically deteriorate. Meanwhile, even if the defender can retrain the model periodically to include new patterns, retraining is not only time/resource-consuming but also potentially harmful to the model's performance on \textit{normal} data.

\section{Mobility}
\label{sec:mobility}
The uniqueness of human mobility to a particular location is a reflection of the culture and values of the people living there \cite{8822969}. Each location has its own characteristics, which influence how people move, the time they spend in a geographical area, and the motivations for their travels. That said, the best way to model human mobility is by using a real dataset obtained from real UE mobility in a live network.

Since there are, to our knowledge, no available UE mobility datasets collected by network operators, we decided to generate a synthetic one using a mobility simulator created by Ericsson \cite{sims-gaia} using a publicly available mobile network topology. This section explains in more detail the simulator, the legitimate mobilities, and the adversarial mobilities developed on top of the simulator. 

\subsection{Mobility Simulator}
The mobility of a UE is sensitive data that has to be treated carefully. Hence, obtaining real data from a mobile operator is difficult due to privacy concerns governed by strict laws, i.e., General Data Protection Regulation (GDPR) in Europe \cite{gdpr}. To overcome this, Ericsson has developed a spatiotemporal mobility simulator, producing the dataset of UE mobility given a set of pre-determined mobility models. In this work, even though the mobility data is based on a pre-determined model, the spatial aspect of the simulator is based on a real-world deployment by Airtel's open-network topology \cite{airtel-open} in India. 

The dataset of UE mobility is a time series. For a particular time (let's call this a timestamp $t$), it manifests a UE connected to a certain eNodeB with a calculated signal strength depending on its distance and the load of the respective eNodeB. The original mobility simulator contains legitimate mobilities in section \ref{mob:legit}, while the extended version contains additional malicious mobilities from section \ref{mob:adv}. While the simulator is not open-source, a subset of the post-processed dataset from our work is available at the following repository \cite{dataset-nwdaf}.

\subsection{Legitimate Mobilities}
\label{mob:legit}
Accurately modeling human mobility patterns is a significant challenge due to their high variability across different locations worldwide. This variability makes it difficult to develop a generalizable approach for predicting movements at a specific time and place. In our study, we focus on a radio network installation in India (and in particular, Airtel \cite{airtel-open}), with legitimate mobilities consisting of people who are working in an office setup (with a particular and consistent location of office and home), people who drive taxis, and local street vendors who move around in a limited particular area. While this is not an exhaustive list, we select representative mobility types that can be formalized and categorized into the same type, allowing us to make meaningful conclusions based on similar movement types.

\begin{figure}[ht]
    \centering
    \includegraphics[width=1\linewidth]{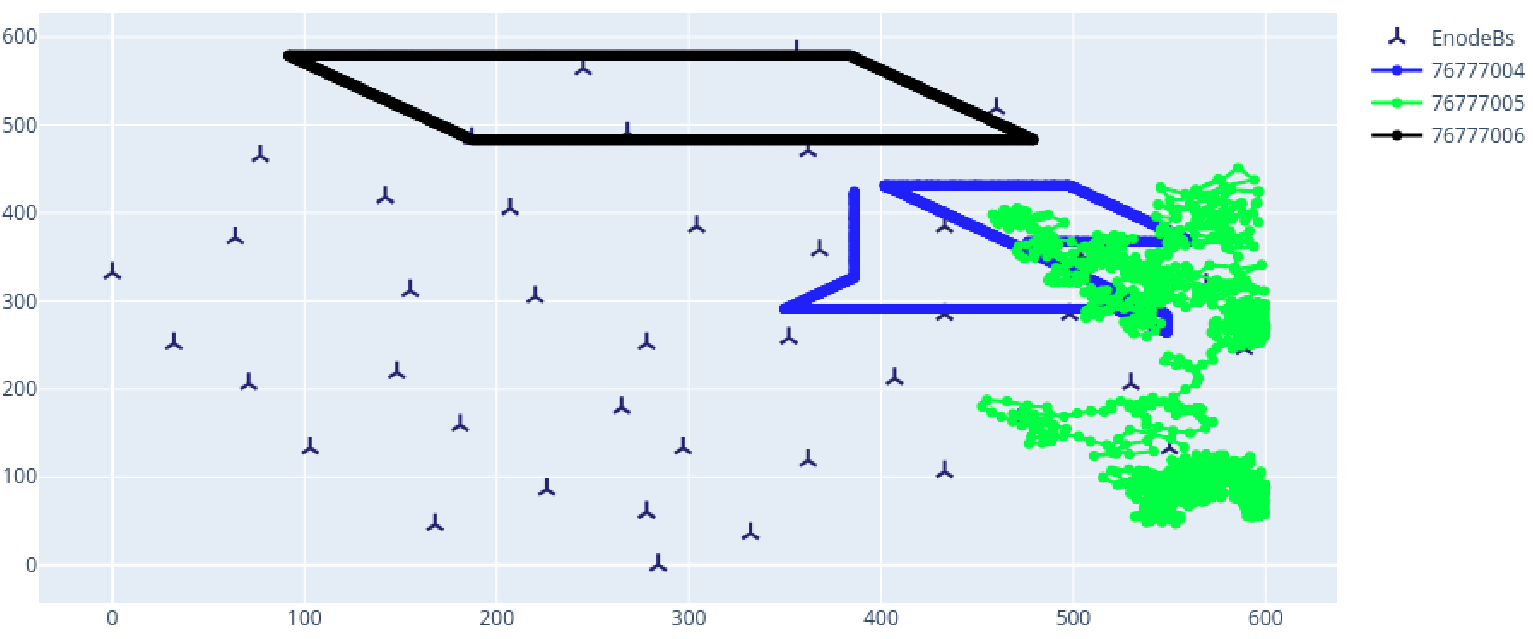}
    \caption{3 UEs with different mobility. Black: Working Professional, Blue: Random Way Point, and Green: Gauss-Markov.}
    \label{fig:sims}
\end{figure}

\subsubsection{Working Professional (WP)}
Working Professional is a custom mobility model designed to simulate the pattern of a UE belonging to an employee with a stationary workplace. It follows a periodic movement pattern between the office and the home. Professionals are assumed to move between home and office during the day and evening. Initial coordinates for home and office are randomly selected (for a given constraint, i.e., specific boundaries on a particular location) when the mobility model is constructed. Lastly, the mobility pattern is assumed to happen on a daily basis. Such mobility has been formalized in \cite{10.1145/1374688.1374695} and represents the population of 140 million people in India\footnote{https://www.thehindu.com/data/India-walks-to-work-Census/article60346511.ece}.

\subsubsection{Random Way Point (RWP)}
The Random Waypoint \cite{hyytia2007random} model simulates random UE mobility behavior over time. It randomly generates destination, speed, and direction for each UE, independently of other UE's, allowing the UE to move freely and without restrictions. This model has been used in numerous simulation studies \cite{9145333,7399689,1498468}. The movement of a UE is governed in the following manner: Each UE may be able to move/pause for a timestep based on a probability value between 0 and 1. At a particular time, this indicates how likely a UE is going to move or not. At each step, a random destination, minimum speed, and maximum speed are selected. The UE moves from its current location to the randomly selected destination and repeats this process for the duration of the simulation. This mobility is analogous to a taxi driver, where the direction and the time spent at each point are uniformly random, depending on the request from the passenger. There are estimated to be around 1.9 million taxi drivers in India\footnote{https://mobilityforesights.com/product/on-demand-taxi-market-in-india/}. While they are still a fraction of the total population, the mobility behavior of taxi drivers is common in downtown areas (regardless of the country) in general. Another reason why we include RWP is that, unlike WP, which usually takes only one round-trip per day, a taxi driver often takes more than a trip, depending on how many customers they serve, generating a larger number of mobility events.

\subsubsection{Gauss-Markov (GM)}
The Gauss-Markov model \cite{4141345} is a mobility model that enables objects to move across a grid plane, with their next transition point determined by their current speed and direction. One can think of this model as similar to a local vendor/salesperson walking around a particular area. There are estimated to be about 10 million street vendors in India\footnote{https://www.indiaspend.com/governance/only-11-of-vulnerable-street-vendors-benefitted-from-pm-credit-scheme-survey-774968}.

Figure \ref{fig:sims} shows the movement of WP (Black), RWP (Blue), and GM (Green) across a particular Airtel India's cartesian plane. For illustrative purposes, each mobility is shown with a single UE, while the actual simulation involved 10,000 UEs for each mobility.

While WP is intuitively the most predictable mobility among the three, the unpredictability of random mobility models such as RWP and GM presents a challenge for accurate prediction, as movement patterns are inherently random. To address this challenge, it may be more important to determine the proportion of unpredictable UEs relative to the total number of mobile subscribers in a given area. This information can be used to allocate appropriate resources for these unpredictable UEs. Despite the seemingly random nature of their movements, it is important to note that random UEs are still subject to the laws of physics, and therefore, it is impossible for a UE to instantaneously jump from one location to another. By accounting for the proportion of random UEs and the physical limitations of their movement, it may be possible to develop more effective strategies for managing network resources in areas with high levels of random mobility.

\subsection{Prediction Accuracy}
In the case of mobility prediction, there are two important metrics to look into:
\begin{enumerate}
    \item \label{corpred}Location prediction at a particular time, denoted as $\hat{l}$.
    \item \label{translot}Timeslot prediction ($\hat{s}$), given that $\hat{l}$ is correct. This shows how long a UE stays in a particular enodeB at a particular time.  
\end{enumerate}

Supposed that there are $n$ mobility events $e$ for a period of $t_2-t_1 \in T$, the mobility prediction accuracy is the sum of correct timeslot prediction $\hat{s}$ (given that the location prediction $\hat{l}$ is also correct), divided by the total number of predictions. Formally, the total accuracy can be written as follows:

\begin{equation}
\label{eq:accuracy}
\textbf{Accuracy} = \frac{\sum_{e=0}^{n} (\hat{s}_{correct} | \hat{l}_{correct})}{n}
\end{equation}

This accuracy calculation from equation \ref{eq:accuracy} will be used for the rest of this paper. 

\subsection{Adversarial Mobility}
\label{mob:adv}

\begin{figure}[ht]
    \centering
    \includegraphics[width=1\linewidth]{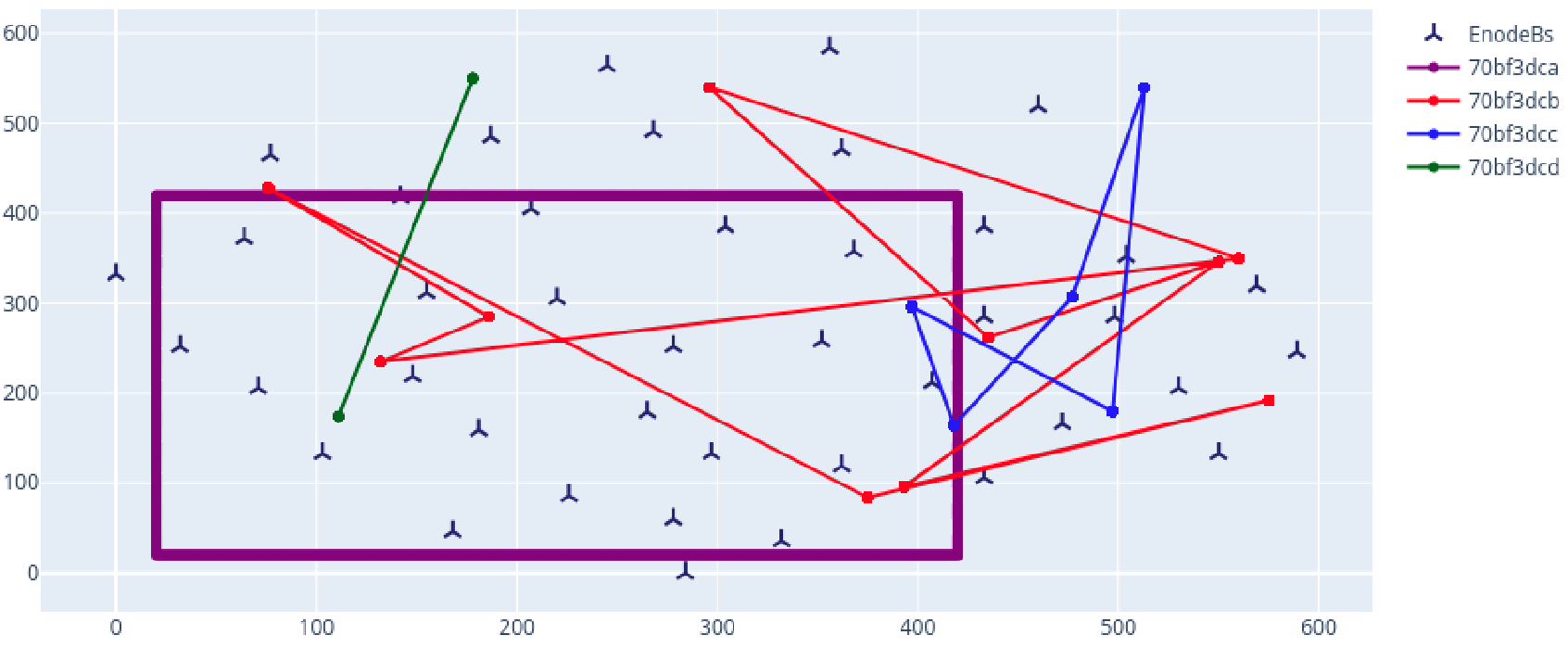}
    \caption{3 UEs with different attack patterns. Green: Tuple Jump, Blue: Quintuple Jump, Red: Decuple Jump, and Purple: Google Maps Attack.}
    \label{fig:simsattack}
\end{figure}

This section outlines the existence of adversarial UEs with the intention of reducing the accuracy of legitimate mobility patterns by introducing adversarial movements. Thus, shifting the perceived UE's behavior such that the operator feels the need to retrain the NWDAF's model with newer data (including adversarial UEs). The adversarial movements are required to be economically efficient by committing to the lowest possible resources while aiming for the highest decrease in accuracy. The resources available to the adversary include the number of acquired UEs and the performed adversarial mobility. The best-case scenario for the adversary is to produce adversarial movement without physically carrying around the UEs, as it minimizes the effort required. However, if the UEs have to be physically transported, the effort involved must be minimized to ensure the sustainability of the attack. In this scenario, an adversary, denoted as $a$, successfully acquires $n$ UEs, and we aim to understand the optimal strategy for $a$ to achieve the lowest accuracy as quickly as possible using the lowest resources in hand.

\subsubsection{Tuple Jump Mobility Attack}
Given that $a$ acquired $n$ UEs, $a$ divides $n$ into $n/2$ paired sets. A set contains two distinctively different UEs but has the same identity, i.e., International Mobile Subscriber Identity (IMSI). As the basis of our assumption, it has been proven that it is feasible to clone the UE, as explained in our threat model. We built the adversarial mobility on top of this argument. 

Within a set, let's call the first UE $u_1$ and the second one $u_2$. Even though they are distinctively different, from the network operator standpoint, $u_1$ and $u_2$ are the same (due to the same identifier, IMSI). The following sequences illustrate the attack:
\begin{enumerate}
	\item At timestamp $t_1$, $u_1$ connects to an enodeB $e_1$. At the same time, $u_2$ stays silent and does not perform any effort to connect to any enodeB.
	\item At timestamp $t_2$, $u_1$ shuts down its interface, while at the same time, $u_2$ connects to another enodeB $e_2$ located on a different geographical location than $e_1$.
	\item Repeat from the $1^{st}$ step. 
\end{enumerate}

What the network operator sees is a UE jumping around from $e_1$ to $e_2$ and back forever. The preference on which $e_1$ and $e_2$ should be connected is uniformly random, selected by adversaries. Not only $e_{1-2}$ can be located adjacent to each other but also in a sparsely different location where movement from $e_1$ to $e_2$ can be impossible from the law of physics. 

However, it is limited to the cartesian plane that we extract from Airtel's open-network \cite{airtel-open}. Using this approach, $a$ does not have to carry around $u_1$ and $u_2$ to make both of them look like they are moving from $e_1$ to $e_2$. The effort to perform the mobility on this attack is virtually non-existent as $u_1$ and $u_2$ can be statically placed near $e_1$ and $e_2$, respectively. We consider the effort to put $u_{1-2}$ in the first place to be negligible, as it is only needed when $a$ deploys the adversarial UEs, not when operating them. 

\subsubsection{Quintuple Jump Mobility Attack}
While in the previous attack model, $a$ divides $n$ into $n/2$, in this scenario, $a$ divides $n$ into $n/5$, expanding the tuple $[u_1, u_2]$ into a quintuple $[u_1, u_2, u_3, u_4, u_5]$. We would like to understand whether expanding the tuple has any positive effect on the decreasing accuracy or not. A set contains UEs with a cloned identity; hence they are perceived to be the same UE by the mobile operator. The sequences are similar. But, instead of repeating in step number 3, the jumping traverses through all the members $u_{1-5}$ and repeats the sequence in step 6. Similar to the previous one, the sets can be vertically scaled up depending on the power of $a$.

\subsubsection{Decuple Jump Mobility Attack}
Similar to the two previous attacks, the number of acquired UEs $n$ are divided into sets, where one set is a decuple, consisting of $[u_{1-10}]$. These sets can be vertically scaled up, depending on the power of $a$.

\subsubsection{Google Maps Attack}
An adversary $a$ acquires a set of $n$ UEs and walks a particular path while carrying them. This type of attack has similarities to a demonstrated attack against Google Maps, where $a$ attempts to create a fake traffic jam in the application by carrying UEs in a basket \cite{gmaps-attack}. We aim to examine whether this attack can be applied to the mobility trajectory in NWDAF. Additionally, we explore the scalability of this attack by varying the number of acquired UEs.

Figure \ref{fig:simsattack} shows the movement of tuple jump, quintuple jump, decuple jump, and google maps attack in green, blue, red, and purple, respectively. Similar to the picture of legitimate mobilities, we only show the movement of each mobility with one UE, spanned over two days of simulation. 

\section{Simulation Environment}
\label{sec:simulation}
The simulation of UE mobility was conducted on a server equipped with 40 Intel Xeon E5-2650 cores and 128 GB memory. In the baseline simulation (i.e., without attack), 10,000 UEs were included in each of the legitimate mobility, WP, RWP, and GM scenarios. The simulation spanned five days, with the first four days allocated for training and the $5^{th}$ day used for testing.

On top of that, the actual model used in this simulation is not part of the NWDAF specification, as the 3GPP standard only specifies interfaces and use cases. That said, the model that we developed here, together with the follow-up attack scenario, is an attack against a mobility model with input and output as if it were deployed in NWDAF.

One might argue that our research should have used real-world data instead of a simulated one. However, obtaining real-world data for the use of NWDAF is not as simple as it sounds. Regulation-wise, the data is considered to be sensitive. While anonymization might help, the operators will not give an interested $3^{rd}$ party access to such data. Our work is an initial effort to open up the research direction in the area despite these limitations. We simulate the legitimate mobilities in a way that it should be enough that there are different mobilities, following human mobilities from the previous research \cite{8822969}. It allows the needed amount of data (in order to get significant and reliable experimental results) to be generated under different mobility patterns, reflecting the reality that mobile data is heterogeneous over time and at different geographical locations. Although the used mobility data is synthetic and will differ somewhat from data of a real deployment, we believe that this does not affect the validity of evaluating the threats. The investigated attacks create mobility patterns that differ from the mobility models, and although results would differ for models built from real-world data, the principle would also be applicable to real scenarios.

\section{Result}
\label{sec:result}
The 3GPP standard does not explicitly specify whether one or more models should be used to perform trajectory prediction. In this study, we utilize a different model for each legitimate mobility type. This approach is motivated by two factors. Firstly, certain mobility types are easier to predict than others, as evidenced by the WP mobility (see table \ref{tab:baseline}), which outperformed the other mobility types. Secondly, combining the datasets into a single training and testing set would not make sense since the poor performance of some mobility types would overshadow the good results of others.

We assume that the mobile operator knows the mobility type of each UE before the adversarial UEs are introduced by the attacker. This can be accomplished by examining the UE's history and attaching a label to each IMSI. Alternatively, unsupervised learning can be used to cluster the results into a labeled dataset with mobility types attached to each UE. While previous studies have proposed spatiotemporal clustering techniques for this purpose \cite{10.1007/978-3-030-70626-5_11}\cite{Kisilevich2010}, our work does not include these techniques.

\subsection{Data Preprocessing}
\label{sec:dataprep}
The dataset used in this study, generated by \cite{sims-gaia}, includes three core features: \{IMSI, enodeb\_path, signal\_strength\}. Each unique IMSI serves as an index, with the historical data of visited eNodeBs, their corresponding timestamps and signal strengths recorded in each data point. Take the following example of a raw data point:

\begin{table}[ht]
\caption{An example of a raw, IMSI-based data point}
\centering
\begin{tabular}{p{0.2\linewidth}p{0.3\linewidth}p{0.3\linewidth}}
\toprule
\textbf{IMSI} & \textbf{enodeb\_path} & \textbf{signal\_strength} \\ 
\midrule
09ccc864 & {'163':12,'348':10,...} & {'163':0.189,'348':0.186,...} \\
\bottomrule
\end{tabular}
\end{table}

This shows a UE with an IMSI \texttt{09ccc864} that connects to an eNodeB 12 at timestamp 163 with a signal strength of 0.189, followed by connecting to eNodeB 10 at timestamp 348 with a signal strength of 0.186, and so on. To perform a timewise prediction, where the NWDAF predicts which eNodeB a UE will connect to and for how long it will stay connected at a particular time, the dataset is transformed from IMSI-based to an event-based. An event occurs when a UE connects to an eNodeB with one timestamp representing one minute. The duration of time that a UE stays connected to a particular eNodeB is calculated as the time elapsed between the current and the next connection to another eNodeB. This calculation is used to create the timeslot column for each event, providing more complete information. Given the example above, a timeslot with the value of 185 (coming from 348-163) is added to the timeslot column of the data point.

To contextualize the dataset, we incorporate historical information by appending the values of the previous four connected eNodeBs and their respective timestamps (relative to each day) to each data point. Additionally, we include the top four neighboring eNodeBs based on the adjacency matrix of Airtel's eNodeB network installation \cite{airtel-open} to provide an additional location context. This information enables us to determine the top four nearest eNodeBs available for a given target eNodeB location. We categorize the timestamp of each IMSI's mobility into five distinct time bins: \texttt{\{early\_morning, morning, noon, evening, and night\}}, and then aggregate the sum of each bin to provide information on how each IMSI spends its time on a particular day. Finally, we also include the eNodeB to which a UE stay connected the longest, called home\_enodeb. The final input variable (\texttt{$x$}) includes the following enriched features: 

\begin{lstlisting}[breaklines=true]
{enode_1, enode_2, enode_3, enode_4, time_1, time_2, time_3, time_4, target_time, sig_st, imsi, home_enb, early_morn, morning, noon, evening, night, neigh_1, neigh_2, neigh_3, neigh_4}
\end{lstlisting}

The dataset is used to predict a binned timeslot ($\hat{s}$) that a UE stays connected to a particular eNodeB ($\hat{l}$) using the model discussed in Section \ref{sec:baseline}.

\subsection{Baseline}
\label{sec:baseline}

\begin{table}[ht]
\caption{Baseline accuracy for each mobility type}
\centering
\begin{tabularx}{\linewidth}{@{}cXXX@{}}
\toprule
 & \textbf{WP} & \textbf{RWP} & \textbf{GM} \\ 
\midrule
\textbf{Accuracy} & 75.2\% & 44.8\% & 37.6\% \\
\bottomrule
\end{tabularx}
\label{tab:baseline}
\end{table}

Table \ref{tab:baseline} presents the baseline accuracy of each mobility type when there are no adversaries. The results indicate that the mobility type of working professionals has the highest accuracy since it is the most predictable. Conversely, the other mobility types perform poorly due to their randomness in destination, speed, and direction (for the random waypoint) and a random variable sampled from the Gaussian distribution (for the Gaussian-Markov).

Each event previously mentioned in section \ref{sec:dataprep} is independently treated to produce two predictions (namely $\hat{s}$ and $\hat{l}$), i.e., the models have been trained independently of each other. The baseline accuracy presented in Table \ref{tab:baseline} is obtained by running the classifier in Table \ref{tab:model} on the dataset previously pre-processed, with the classifier and its hyperparameters automatically selected by the FLAML AutoML framework \cite{wang2021flaml}. We did not manually choose the classifier or its hyperparameters.

\begin{table}[ht]
\caption{Model used for each mobility type}
\begin{tabularx}{\columnwidth}{>{\hsize=.5\hsize}X>{\hsize=1.25\hsize}X>{\hsize=1.25\hsize}X}
\toprule
\textbf{Mobility} & \textbf{Model $\hat{l}$} & \textbf{Model $\hat{s}$}  \\ 
\midrule
\textbf{WP} & Random Forest \cite{breiman2001random} & Extra Tree \cite{geurts2006extremely} \\
\midrule
\textbf{RWP} & Random Forest \cite{breiman2001random} & Extra Tree \cite{geurts2006extremely} \\
\midrule
\textbf{GM} & Light GBM \cite{NIPS2017_6449f44a} & Extra Tree \cite{geurts2006extremely} \\ 
\bottomrule
\end{tabularx}
\label{tab:model}
\end{table}

\subsection{Adversaries Included}
\begin{figure}[ht]
    \centering
    \includegraphics[width=0.85\linewidth]{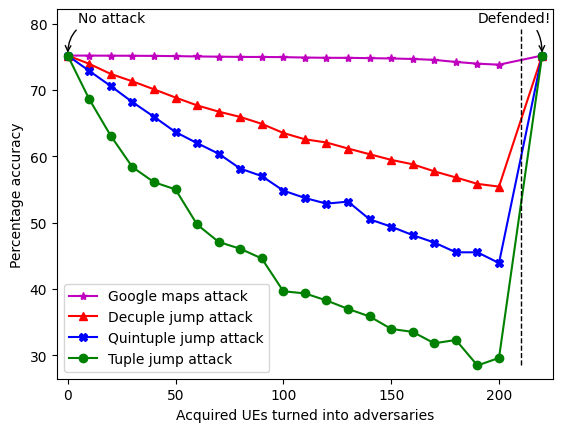}
    \caption{The number of acquired adversarial UEs on the accuracy of WP's mobility}
    \label{fig:wp}
\end{figure}

\begin{figure}[ht]
    \centering
    \includegraphics[width=0.85\linewidth]{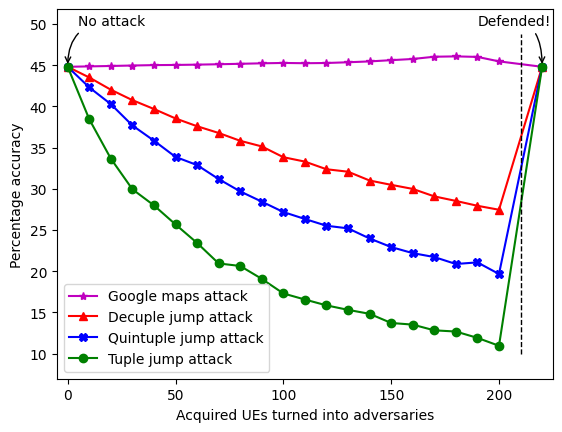}
    \caption{The number of acquired adversarial UEs on the accuracy of RWP's mobility}
    \label{fig:rwp}
\end{figure}

\begin{figure}[ht]
    \centering
    \includegraphics[width=0.85\linewidth]{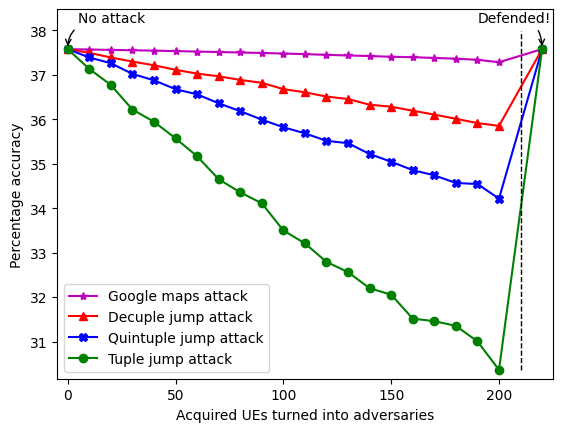}
    \caption{The number of acquired adversarial UEs on the accuracy of GM's mobility}
    \label{fig:gm}
\end{figure}

Figures \ref{fig:wp}, \ref{fig:rwp}, and \ref{fig:gm} demonstrate the impact of adversarial mobility on the total mobility prediction accuracy. Each graph shows the changes in accuracy for the different legitimate mobility types as the number of acquired adversarial UEs increases. The dataset was expanded by the data points coming from the adversarial UEs when the attack occurred. After preprocessing, there were no noticeable changes (decrease or increase) in the accuracy value if we only selected the legitimate data points. However, the accuracy value decreased due to the addition of events with low accuracy from the adversarial UEs. We did not observe any connection between the introduction of the adversarial UEs and the accuracy value of legitimate UEs. This is because the preprocessing mechanism, as explained in section \ref{sec:dataprep}, does not involve mixing inter-IMSI data when we add historical context to each data point. Nevertheless, NWDAF treats all UEs as a legitimate source of data, as it is unaware of the existence of adversarial UEs at this point. In short, by introducing adversarial events, the mean accuracy will go down by the proportion of adversarial events.

When $n$ UEs are acquired, the optimal strategy for the adversary is to execute the tuple jump attack. This involves dividing the attacker UEs into $n/2$ pairs of cloned IMSIs, maximizing the decrease in accuracy. The quintuple and decuple jump attacks also decrease the accuracy number, but not as fast as the tuple jump attack.

The tuple jump attack stands out as the most effective because it generates a greater number of data points compared to other attacks. For instance, given $T \ni t_0-t_{10}$ and 10 acquired UEs by $a$, at $t_{10}$, the generated data points are $\{50,20,10\}$ for \{tuple, quintuple, decuple\}. Moreover, at any timestamp, the tuple jump attack has only one idle UE, while the quintuple and decuple jump attacks have four and nine idle UEs, respectively. While having the idle UE is necessary due to the unwillingness of the adversary to physically carry them around, optimizing their existence is necessary, and one way to do that is by minimizing the number of idle UEs.

Although the various jump attacks have different effects on accuracy, the Google Maps attack appears to have little impact on overall accuracy. Graphs for WP and GM indicate a slight decrease in accuracy as the number of acquired UEs increases, but the opposite occurs for RWP, where accuracy actually slightly improves. One possible explanation for this phenomenon could be attributed to the intrinsic randomness of the RWP mobility model, which makes it challenging to develop an accurate prediction model. In contrast, the Google Maps attack involved a set of devices with more predictable movements. Strikingly, as the proportion of predictable devices increased in the dataset, we observed a paradoxical effect: the accuracy of the prediction model improved instead of decreasing, as one might expect. Therefore, an adversary seeking to quickly decrease accuracy would not need to use this method, at least not against an NWDAF model with the UEs comprised of WP, RWP, and GM mobility.

\section{Defense Mechanism of The Deployed Model}
\label{sec:def}
In light of the attacks discussed in the previous section, safeguarding the trustworthiness of legitimate users is paramount. Although the accuracy of legitimate UE may remain unchanged, it is imperative to defend against a decrease in total accuracy (legitimate+adversarial) since a general mobility prediction model is unable to distinguish adversarial data.

As a way to analyze the properties of the adversarial data, i.e., how easy it is to distinguish the adversarial from the legitimate data, we employ KMeans \cite{hartigan1979algorithm}. It is chosen due to its simplicity and the most popular clustering methods. The adversarial samples in our dataset are distinct from normal samples, making them easy to cluster using KMeans. While KMeans is not intended for real-time detection of adversarial UEs, it can be used to solve the problem of malicious mobility patterns by identifying which data is safe or unsafe to be included in the retraining process. This is because the main problem, as explained in Section \ref{threadmodel}, is that a shift in UE behavior can lead the mobile operator to believe that retraining with newer data is necessary.

Although there may be other methods to defend against adversarial UEs, such as increasing the robustness of the deployed model, we leave this as an open question. We believe that more attacks will be introduced in the future, and the solution will depend on the respective attacks. However, for adversaries that fall within the constraints defined in our threat model, KMeans serves as an adequate distinguisher.

Given a set of UE location data $\mathbf{X} = \{\mathbf{x}_1, ..., \mathbf{x}_n\}$, $x_N \in \mathbf{R}^d$, where $\mathbf{x}_i$ has three features, i.e., \{IMSI, enodeb\_path, signal\_strength\}, KMeans splits the dataset that includes both malicious and benign samples into $K$ disjoint sub-clusters $C_1, ..., C_K$. It aims to minimize the distance between a data point and the centroid of every sub-cluster. That is to say, the data sample belongs to the closest sub-cluster. Then, we calculate the hamming distance of the accuracy using each of the clustered labels. Equation \ref{eq:k-meansDistance} is the clustering loss where $f_i$ is the clustering loss, $x_i$ is the $i^{th}$ sample and $o$ is the cluster center. $d(x)$ is the hamming distance function, $N$ is the total number of samples in a dataset. In our settings, we set $K=2$ for malicious and benign clusters.

\begin{equation}
    f_i = \sum_{k=1}^K \sum_{i=1}^N  g(x_i \in C_K) [d(x_i, o)]^2,
    \label{eq:k-meansDistance}
\end{equation}

\subsection{Dataset}
To evaluate the effectiveness of KMeans, we analyze its performance using three different datasets that pertain to WP, RWP, and GM mobility patterns, where the WP, RWP, and GM datasets contain 110811, 80930, and 613306 normal data points generated by 10000 UEs. Besides, the WP, RWP, and GM datasets include 233949 adversarial data samples, i.e., 142560, 56736, 27792, and 6861 from Tuple Jump, Quintuple Jump, Decuple Jump, and Google Maps attacks. The data points of each attack model are produced by 200 UEs (the highest number of acquired UEs from figures \ref{fig:wp}, \ref{fig:rwp}, and \ref{fig:gm}). In total, there are 805047 legitimate and 701847 malicious samples in our dataset. In our experiments, the parameter n\_clusters=2 as we have two classes, malicious and benign; as for other parameters, we use the defaults of the KMeans() function in \textit{sklearn}, e.g., random\_state=0, n\_init=10.

\subsection{Results}
\begin{figure}[ht]
    \centering
    \includegraphics[width=.85\linewidth]{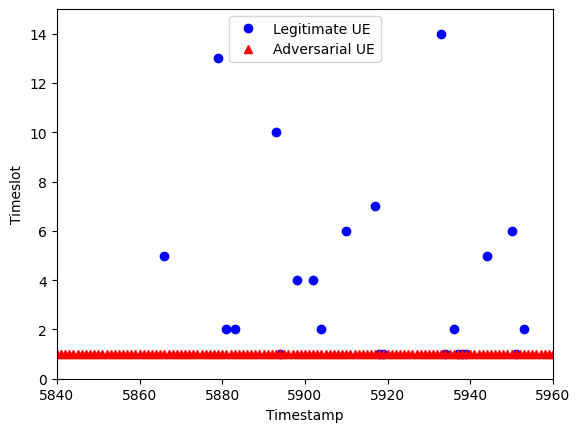}
    \caption{KMeans visualization of a selected timestamp on Gauss-Markov Mobility}
    \label{fig:kmeans}
\end{figure}

The visualization result of KMeans clustering for GM's mobility and the Tuple Jump mobility attack is presented in Figure \ref{fig:kmeans}. The figure displays the timestamps and timeslots of one legitimate and one adversarial UE from a specific timestamp range on the $5^{th}$ day of the simulation. The timeslot is defined as the time duration of a UE staying connected to a specific eNodeB, which is calculated by the elapsed time between the current and the next connection of the UE to another eNodeB. The clustering result demonstrates that KMeans successfully distinguishes between the timeslots of legitimate and adversarial UEs. The legitimate UE has a varied timeslot for each connection, while the Tuple Jump attack has a consistent timeslot throughout the simulation. Although an attacker may randomize the timeslot value to evade the clustering, a lower timeslot means more data points on the attack dataset, which will result in a much faster accuracy reduction. Thus, a more sophisticated adversary must strike a balance between having as many data points as possible and randomizing the timeslot value. We left this as an open question for future research. 

Figures \ref{fig:wp}, \ref{fig:rwp}, and \ref{fig:gm} demonstrate the activation of the defense mechanism at their respective rightmost ends. When the defense mechanism is activated, the adversarial UEs can be separated from the rest of the dataset. It deters the mobile operator from retraining the model with newer data, as this could adversely affect the model's performance. The operator can then take follow-up actions on the separated dataset of adversarial UEs, such as blocking them or taking other measures to mitigate the attack. However, the specific follow-up actions are out of the scope of this paper.

\section{Related Work}
\label{sec:related}
NWDAF is a critical component in 5G networks that enables network automation, optimization, and security, and there has been a growing interest in using NWDAF to enhance data analytics capabilities and enable more advanced automation and optimization of 5G networks. To enhance NWDAF's performance, numerous research studies have been carried out. They can be grouped into three categories, i.e., ``NWDAF-oriented mobility trajectory prediction'', ``adversarial attack on network mobility'', and ``attack defense mechanisms''. 

\textbf{NWDAF-oriented mobility trajectory prediction:} 
In \cite{9730290}, the authors emphasized the significance of NWDAF for enabling real-time data analytics and decision-making in mobile networks and explained that NWDAF plays a crucial role in collecting, processing, and analyzing network data. \cite{init_nwdaf} delves into the integration and implementation of NWDAF in 5G/6G core networks. They discuss the architecture of NWDAF and its capability to provide slice-specific network data analytics to various NFs. Through a case study, the paper illustrates the insights obtained from NWDAF-collected data and highlights its potential in steering traffic policies and resource allocation, including from the mobility data.

Similarly, \cite{shafin2020artificial} highlighted the importance of utilizing machine learning algorithms in NWDAF for beamformer design and predicting mobility for better handover management. Mobility in the 5G core network plays a pivotal role in ensuring efficient and agile operations, as explained in \cite{jeong2021mobility}. The use of machine learning has shown potential in accurately predicting these mobility patterns within the core network domain. However, a significant challenge emerges in the practical application of these predictions. One notable limitation to consider is the sensitivity of the prediction models to changes in the network topology; for instance, the entire Recurrent Neural Network (RNN) model might require retraining if there is an addition or removal of a gNB. Despite these challenges, advanced predictions offer practical advantages: if the network can anticipate a UE move to a target edge site, it allows for preemptive actions in the relocation procedure, streamlining the process before the actual gNB handover occurs.

\textbf{Adversarial attack on network mobility:}
As aforementioned, NWDAF is critical for automating data collection and analytics of the core network, and ML methods play a key role in NWDAF. However, due to the distributed and API-based structure, the ML algorithms in NWDAF could be possible targets for adversarial attacks \cite{ashraf2022zero}. Attacks on the network mobility can occur either during the training or testing phases. Training phase attacks, such as poisoning attacks, aim to corrupt the model by injecting manipulation into the training process that causes the trained model to make incorrect or biased predictions. For example, \cite{zolotukhin2023attacking} considered a poisoning attack in 5G, where compromised UEs are utilized to transfer malicious data to RAN after using their credentials to authenticate and be authorized in the core network. Testing phase attacks attempt to create input data that appears normal but is specifically designed to mislead the machine learning model to give the wrong results \cite{42503}. Our work falls into the category of testing phase attacks, specifically targeting the mobility trajectory model after it is deployed. 

\textbf{Attack defense mechanisms:}
There are mainly two types of defense mechanisms, proactive and reactive. Proactive defenders prevent adversarial attacks before they happen by, for instance, Moving Target Defense (MTD) \cite{soussi2021moving} and distillation \cite{7546524,grosse2017adversarial} \& adversarial training \cite{anthi2021adversarial}, while reactive defenders respond to adversarial attacks when they occur, for example, using anomaly detection \cite{sternby2020anomaly,MENSI2023109334,4781136}. For proactive defenses, MTD \cite{soussi2021moving} defends against the attack by continuously changing networks' attack surface to make it more difficult for attackers to identify and exploit vulnerabilities. By using MTD, it is more challenging for attackers (by reducing their asymmetrical advantage) to find the vulnerabilities of the network, while the network defender can respond to attacks in real-time, and the network can adapt to changing threats. 

Distillation as a defense mechanism \cite{7546524,grosse2017adversarial} involves training both a ``teacher'' and a ``student'' model. The ``teacher'' produces softened output probabilities, which act as labels when training the ``student''. The idea is to have the ``student'' model learn the underlying patterns in the data rather than merely memorizing the training set, which can make it more robust to adversarial attacks. While distillation-based defenses can enhance resilience against certain adversarial attacks by promoting more robust feature learning, they are not universally foolproof \cite{7958570}. In terms of computational efficiency, once the teacher model is trained, only the student model needs additional training, which can be more efficient than other defense methods that require multiple retrainings. Adversarial training is used to improve the robustness of a model against adversarial attacks by incorporating malicious data into the training set to improve the robustness of the model against future attacks. Yet, the computational cost could be increasingly high as it requires constantly generating and including adversarial samples for training. For instance, \cite{anthi2021adversarial} evaluates the performance of a supervised-learning-based industrial control system when it is under adversarial attacks and utilizes the generated adversarial samples to improve the robustness of the system.

For reactive defenses, anomaly detection distinguishes the benign and malicious data. Drawing inspiration from the Isolation Forest algorithm \cite{4781136}, which is a previous method for outlier detection, \cite{MENSI2023109334} extends its concept to a proximity-based framework. Unlike the Isolation Forest, which operates on feature-based outlier detection, the Proximity Isolation Forest requires only a set of pairwise distances to function, making it adaptable to different types of data. Anomaly Detection Forest \cite{sternby2020anomaly} discusses the challenge of anomaly detection in a case where labeled anomaly data is not available and proposes a new algorithm designed for one-class learning with only normal instances as training data. The algorithm is an ensemble of binary trees and shows superiority over existing algorithms, i.e., Isolation Forest and One-Class Random Forest for one-class anomaly detection. While the purpose is not to perform a real-time classification, our work in this paper is considered a reactive defense, where we employ KMeans to distinguish the adversarial UEs from the legitimate ones. 

\section{Conclusion}
\label{sec:conclusion}
This study presents an investigation of potential attacks against mobility trajectory models supporting 3GPP NWDAF abnormal mobility normative analytics use cases. The results show that, under the constraint of an adversary who aims to minimize their effort along with the power of IMSI cloning, a tuple jump attack is the most effective strategy. This attack involves dividing the acquired adversarial UEs into pairs of cloned IMSIs, resulting in a decrease in mobility prediction accuracy from 75\% to 40\%, given 10,000 legitimate subscribers and 100 acquired adversarial UEs. The purpose of such an attack is to make the mobile operator believe that there is a shift in the UE's mobility behavior, and retraining with newer data is necessary, which leads to a lower accuracy on the legitimate UEs if this happens. While our attack works on a given legitimate mobility, i.e., working professional, random-waypoint, and gaussian-markov, further study is needed for different mobility types. Our analysis shows that the data from adversarial movements is distinct, and KMeans Lloyd's unsupervised clustering technique is sufficient to distinguish between legitimate and adversarial movements.

Future research directions include i.) testing the approach on a live network dataset and investigating the implications of retraining the model with adversarial data, ii.) modeling more sophisticated adversarial UEs, i.e., armed with the power to randomize the timeslot when jumping between points in the cartesian plane. This would need a more advanced defense strategy, as we would expect. And lastly, iii.) Expanding the investigation to include cellular IoT scenarios, which are expected to dominate the amount of 5G subscriptions due to the support of massive IoT deployments in 5G.

The findings of this study provide insights for the development of more robust NWDAF-supported mobility prediction ML models that can effectively defend against such attacks.

\section*{Acknowledgment}
\label{sec:ack}
We would like to thank Michael Liljenstam and Raaghul Ranganathan for their insightful feedback on the initial version of this paper.

\bibliographystyle{IEEEtran}
\bibliography{references}
\end{document}